\documentclass[sigconf,9pt,screen=true,anonymous=false,bookmarks=false,nonacm]{acmart}

\usepackage{lipsum}
\usepackage{graphicx}
\usepackage{footnote}

\usepackage{amssymb}

\usepackage{footnote}
\usepackage{amsmath}
\usepackage{mathtools}
\usepackage{mathrsfs}                                      
\usepackage{siunitx} 
\usepackage{comment}
\usepackage{wrapfig} 
\usepackage{hyperref}
\usepackage[subrefformat=parens,labelformat=parens]{subfig}
\usepackage{bm,bbm}
\usepackage{multirow}
\usepackage{makecell}
\usepackage{threeparttable,booktabs}
\usepackage{blkarray}
\usepackage{tikz}
\usetikzlibrary{positioning,calc,fit,decorations.pathmorphing,shapes.geometric,shapes.gates.logic.US,calc}
\usepackage{tikz-3dplot}
\usepackage{balance}
\usepackage{courier}                                       
\usepackage{pifont}                                        
\usepackage{cleveref}                                      
\usepackage[mathcal]{eucal}
\usepackage[]{algpseudocode}                               
\algrenewcommand\textproc{\texttt}
\makeatletter\let\float@addtolists\relax\makeatother
\usepackage{algorithm}
\algblockdefx[pardo]{pardo}{endpardo}%
   [1]{\textbf{For} #1 \textbf{do in parallel}}%
   {\textbf{Endfor}}
\usepackage{filecontents}                                  
\usepackage{pgfplots}
\usepackage{pgfplotstable}
\usepackage{pgf-pie}
\usepgfplotslibrary{groupplots}
\pgfplotsset{compat=newest}
\usepackage[figuresright]{rotating}
\usepackage{xcolor,colortbl}                               
\usepackage[inline]{enumitem}                              
\setlist{leftmargin=5.08mm}

\newcommand{\minisection}[1]{\vspace{.06in}\noindent{\textbf{#1}}.}

\DeclareMathOperator*{\argmax}{argmax}

\theoremstyle{plain}

\theoremstyle{definition}

\algrenewcommand\textproc{\texttt}

\usepackage[skip=1pt]{caption}            
\definecolor{CUHKorange}{RGB}{244,106,18} 
\definecolor{CUHKblue}{RGB}{0,111,190}    
\definecolor{CUHKgreen}{RGB}{0,127,128}   
\definecolor{CUHKred}{RGB}{228,46,36}     
\definecolor{CUHKyellow}{RGB}{198,148,34} 
\definecolor{CUHKdark}{RGB}{114,44,114}   
\definecolor{CUHKmiddle}{RGB}{144,44,144} 

\usepackage{tcolorbox}
\tcbuselibrary{skins,breakable}
    {\endtcolorbox}

\graphicspath{{./figs/}}

\copyrightyear{2024} 
\acmYear{2024} 
\setcopyright{rightsretained} 
\acmConference[ICCAD '24]{IEEE/ACM International Conference on
Computer-Aided Design}{October 27--31, 2024}{New York, NY, USA}
\acmBooktitle{IEEE/ACM International Conference on Computer-Aided Design
(ICCAD '24), October 27--31, 2024, New York, NY, USA}
\acmDOI{10.1145/3676536.3676775}
\acmISBN{979-8-4007-1077-3/24/10}

\usepackage{threeparttable}
\usepackage{makecell}

\usepackage{listings}

\usepackage{xcolor} 
\definecolor{codegreen}{rgb}{0,0.6,0} 

\usepackage{fancybox}  

\newif\ifrev
\revtrue
\ifrev
  \newcommand{\bei}[1]{{\color{red} [Bei: #1]}} 
\else
  \newcommand{\bei}[1]{}
\fi

\definecolor{myblue}{RGB}{29,114,221}    
\definecolor{myyellow}{RGB}{255,255,191} 
\definecolor{myorange}{RGB}{244,106,18}  
\definecolor{mygray}{RGB}{102,102,102}   
\definecolor{mypink}{RGB}{252,228,215}   

\definecolor{CUpurple}{RGB}{136,43,142}
\definecolor{CUlpurple}{RGB}{165,133,182}
\definecolor{CUgold}{RGB}{221,163,0}
\definecolor{CUribbon}{RGB}{244,223,176}
\definecolor{CUblack}{RGB}{34,24,21}
\definecolor{PKUred}{RGB}{126,24,28}
\definecolor{gray6}{gray}{0.6}
\definecolor{gray7}{gray}{0.7}
\definecolor{gray8}{gray}{0.8}
\definecolor{gray9}{gray}{0.9}

\usepackage{color}

\begin{document}
\date{}

\title{
  RTLRewriter: Methodologies for Large Models aided RTL Code Optimization
}

\author{
    Xufeng Yao$^1$, \quad
    Yiwen Wang$^2$, \quad
    Xing Li$^2$, \quad
    Yingzhao Lian$^2$, \quad
    Ran Chen$^2$, \quad
    Lei Chen$^2$, \quad
    Mingxuan Yuan$^2$, \quad
    Hong Xu$^1$, \quad
    Bei Yu$^1$ \\
    $^1$Chinese University of Hong Kong \quad $^2$ Huawei  \\
}

\thanks{Xufeng Yao and Yiwen Wang are equally contributed.}

\begin{abstract}
Register Transfer Level (RTL) code optimization is crucial for enhancing the efficiency and performance of digital circuits during early synthesis stages. 
Currently, optimization relies heavily on manual efforts by skilled engineers, often requiring multiple iterations based on synthesis feedback. In contrast, existing compiler-based methods fall short in addressing complex designs.
This paper introduces RTLRewriter, an innovative framework that leverages large models to optimize RTL code.
A circuit partition pipeline is utilized for fast synthesis and efficient rewriting.
A multi-modal program analysis is proposed to
incorporate vital visual diagram information as optimization cues.
A specialized search engine is designed to identify useful optimization guides, algorithms, and code snippets that enhance the model’s ability to generate optimized RTL.
Additionally, we introduce a Cost-aware Monte Carlo Tree Search (C-MCTS) algorithm for efficient rewriting, managing diverse retrieved contents and steering the rewriting results.
Furthermore, a fast verification pipeline is proposed to reduce verification cost. 
To cater to the needs of both industry and academia, we propose two benchmarking suites: the Large Rewriter Benchmark, targeting complex scenarios with extensive circuit partitioning, optimization trade-offs, and verification challenges, and the Small Rewriter Benchmark, designed for a wider range of scenarios and patterns. Our comparative analysis with established compilers such as Yosys and E-graph demonstrates significant improvements, highlighting the benefits of integrating large models into the early stages of circuit design.
We provide our benchmarks at \href{https://github.com/yaoxufeng/RTLRewriter-Bench}{https://github.com/yaoxufeng/RTLRewriter-Bench}.

\end{abstract}

\maketitle
\pagestyle{empty}

\section{Introduction}\label{intro}
\label{sec:intro}
\begin{figure}[tb!]
    \raggedleft    \includegraphics[width=.96\linewidth]{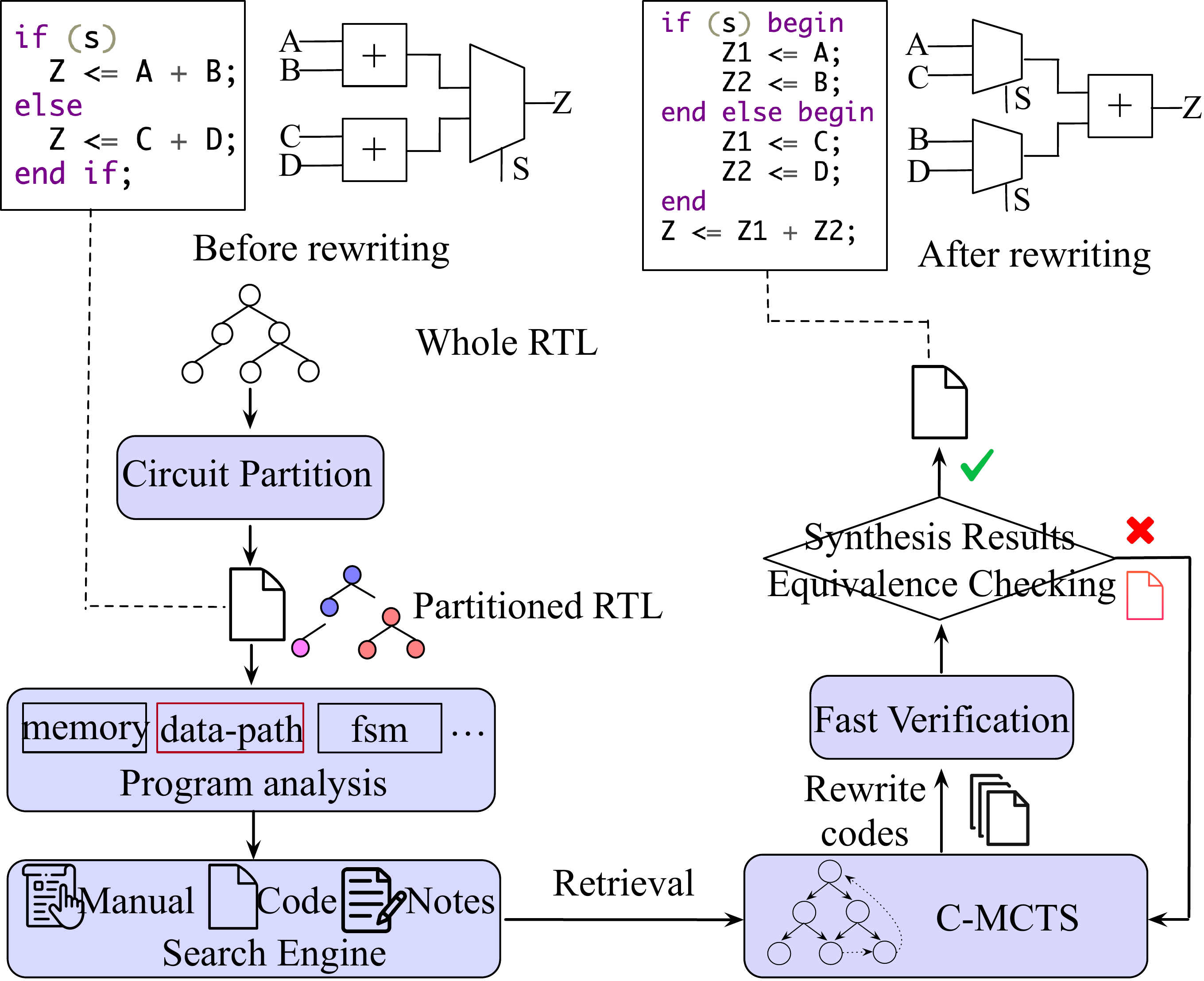} 
    \caption{Rewriting example and Rewriter Pipeline.}
    \label{fig:motivation-pipeline}
\end{figure}

Optimizing Register Transfer Level (RTL) code is an essential step in the early stages of circuit design. 
This process involves multiple rounds of rewriting original RTL code snippets into optimized versions based on optimization patterns or synthesis feedback.
Conventionally, this process relies heavily on the expertise of seasoned engineers. However, the growing complexity of design patterns has significantly hindered the efficiency of manual optimization. 
In comparison, existing compiler-based methods exhibit limited scope and effectiveness in optimizing complex designs, and fall short in optimizing code via synthesis feedback.
~\Cref{fig:motivation-pipeline} illustrates a classic example of MUX optimization, where the revised version achieves a reduction in area by eliminating an adder. Nonetheless, certain open-source compilers, such as Yosys~\cite{wolf2013yosys}, struggle to effectively manage such scenarios.

Previous works on RTL code optimization mainly focus on specific scenarios such as data-path, MUX, memory~\cite{pasko1999new,wegman1991constant,chen2004register,pivstekareduction,wang2023optimization}.
Nevertheless, the real circuit design is more complicated, which contains many optimization trade-off that can not be easily addressed by compilers.
In contrast, skilled engineers are good at rewriting the RTL code via synthesis feedback, taking manual as reference and directing the optimization process efficiently based on their experiences.

To bridge this gap, we introduce RTLRewriter, a framework that leveraging large models for \textbf{whole RTL Code optimization process}. 
~\Cref{fig:motivation-pipeline} illustrates proposed RTLRewriter pipeline.
We have developed multiple components within the framework to address various challenges.
\begin{itemize}
    [leftmargin=12pt]
    \item \textbf{Circuit Partition:} Large models often struggle with long contexts, which is a significant challenge in RTL code optimization. To mitigate this, we implement a circuit partitioning pipeline that breaks down the entire circuit into smaller, manageable segments, enabling faster synthesis and more effective rewriting.
    
    \item \textbf{Semantic Extraction:} Current compiler-based methods fall short in extracting meaningful semantic information from the code, particularly in terms of visual information. We address this limitation by proposing a multi-modal program analysis technique that enhances semantic understanding. 
    
    \item \textbf{Documentation Utilization:} Traditional compilers generally underutilize extensive documentation, notes, and historical code. We propose a dedicated search engine to retrieve relevant content that assists in large model generation.
    
    \item \textbf{Cost-Effective Rewriting:}     Not all retrieved content is beneficial for model generation, and optimal rewriting often requires multiple iterations with suitable prompts. Given the high inference costs associated with large models, we introduce a Cost-aware Monte Carlo Tree Search (C-MCTS) algorithm to efficiently determine the best rewriting strategies.
    
    \item \textbf{Verification Cost Reduction:} To reduce verification costs, we leverage the program analysis capabilities of large models to select the most appropriate solver for verification process.
\end{itemize}
These five components form the core of our framework. When given a complete circuit, we first partition it into smaller parts. Next, we perform program analysis using multi-modal information, generating associated optimization patterns and verification patterns.
Using various elements, such as code and related diagrams, our search engine retrieves relevant content, providing optimization cues for large model rewriting. We then use the C-MCTS algorithm for efficient rewriting.
To address the randomness of large model generation, our fast verification pipeline filters out incorrect instances, selecting the most suitable solver for verification based on program analysis. If the instances pass equivalence checking and yield improved synthesis results, the code is successfully rewritten.

To meet the requirements of both industry and academia, we put forth two RTL rewriting benchmarks in this study.
These benchmarks encompass various examples, consisting of the original RTL code and its corresponding optimized version.
The long benchmark comprises several extensive RTL cases, such as CPU designs and neural networks, presenting significant challenges akin to real-world industry scenarios.
In contrast, the Small benchmark encompasses multiple small RTL cases with different scenarios, each exhibiting distinct optimization patterns.
Further descriptions and details regarding these benchmarks will be provided in the experiments section.

Our major contributions are summarized as follows.
\begin{itemize}
    [leftmargin=20pt]
    \item We introduce the first LLM-aided RTL code optimization framework, leveraging the capabilities of large language models for RTL code optimization.
    \item Our framework incorporates multiple components to tackle various challenges.
    \item We propose two RTL code optimization benchmarks, curated by senior Verilog engineers.
    \item With experiments on different benchmarks, we proved our framework can significantly improve the synthesis performance compared with other LLM-based baselines and competitive compilers such as Yosys and E-graph.
\end{itemize}

\section{Related Works}

\subsection{RTL Code Optimization} \label{ssec: rtl-opt}
RTL code optimization is a long-standing problem in VLSI that has a significant impact on final PPA (Power, Performance, Area). 
While previous works have made substantial efforts toward RTL optimization, thoroughly optimizing RTL code remains challenging. In real industry applications, RTL code optimization heavily relies on experienced Verilog engineers, often requiring multiple iterations of modification based on synthesis feedback.
In the realm of data-path optimization, our focus primarily lies on techniques such as  subexpression elimination \cite{cocke1970global, pasko1999new}, Constant folding~\cite{cocke1970global}, Constant propagation~\cite{wegman1991constant, metzger1993interprocedural}, Algebraic simplification~\cite{buchberger1982algebraic, carette2004understanding}, Dead code elimination \cite{knoop1994partial, gupta1997path}, and Strength reduction \cite{cooper2001operator}.
When it comes to Mux optimization, our considerations revolve around Mux reduction \cite{chen2004register, wang2023optimization}, Mux tree decomposition \cite{pivstekareduction}, and Mux tree restructuring \cite{wang2023optimization}.
In regard to memory optimization, we delve into various areas such as memory sharing \cite{laforest2010efficient, ma2020hypervisor}, memory folding, memory banking \cite{zhou2017new,lai2019remap+} and memory pipelining \cite{park2001synthesis} into consideration.
Lastly, for FSM design, we consider techniques such as state minimization \cite{kam2013synthesis}, state assignment \cite{villa2012synthesis}, and state decomposition \cite{shelar1999decomposition}.
In contrast, compiler-based methods fall short in modifying code via synthesis feedback and are limited in handling complex patterns. To address this gap, we introduce a large model-aided RTL code optimization approach, aiming to provide new directions for automatic RTL code optimization.

\subsection{Large Models aided design
and Challenges}

The utilization of large models into electronic design automation (EDA) is an emerging field, as evidenced by recent advancements in hardware design~\cite{liu2023chipnemo,blocklove2023chip,fu2023gpt4aigchip,zhang2024data4aigchip}, EDA script generation~\cite{wu2024chateda}, RTL code generation and debugging~\cite{thakur2023verigen,liu2023rtlcoder,delorenzo2024make,tsai2023rtlfixer,pei2024betterv,yao2024hdldebugger}, and other applications~\cite{wang2023circuit,wang2024hierarchical,liu2024unified,li2024logic,li2024circuit} These developments highlight the potential of large models to enhance and accelerate circuit design processes.
However, the area of RTL code optimization, characterized by complex patterns and significant verification challenges, remains underexplored. This paper introduces a novel framework that leverages large models to address these issues, effectively filling this research gap.

Additionally, the evaluation of large models in practical industry applications presents new challenges.
For instance, the Verilog generation benchmark, VerilogEval~\cite{liu2023verilogeval}, uses overly simplistic cases from HDLbits. To address this, we introduce two new RTL code optimization benchmarks that include a wider range of case complexities, designed to meet the needs of both academia and industry.

\section{Methodologies }
In this section, we provide a comprehensive overview of the proposed RTLRewriter framework. The first step in our approach is partitioning the RTL design, such as a CPU, into smaller, manageable parts, as detailed in Section~\ref{ssec:parition}. For each sub-circuit, we introduce a novel RTL rewriting framework aided by large models. This framework integrates multi-modal program analysis, a search engine, and a cost-aware Monte Carlo Tree Search algorithm, as outlined in Section~\ref{ssec:rtl-rewriting}. To mitigate the inherent randomness of LLM generation and reduce verification costs, we implement a fast verification pipeline that filters out ineffective rewrites using automated test case generation and efficient program analysis-based verification, as described in Section~\ref{ssec:fast-verification}.

\subsection{Circuit Partition} \label{ssec:parition}

The challenge of comprehending long contexts remains an unsolved problem in transformer-based large models ~\cite{liu2024lost} due to their quadratic complexity, and we have observed this phenomenon in RTL code optimization as well. 
Specifically, we have observed that the effects of rewriting are significantly improved when applied to partitioned sub-circuits, as opposed to the entire circuit.
Furthermore, partitioning a large circuit is a common practice in the industry due to the high synthesis time cost. It involves dividing a large circuit into smaller parts that can be synthesized in parallel.

\begin{figure}[tb!]
    \centering    \includegraphics[width=.9\linewidth]{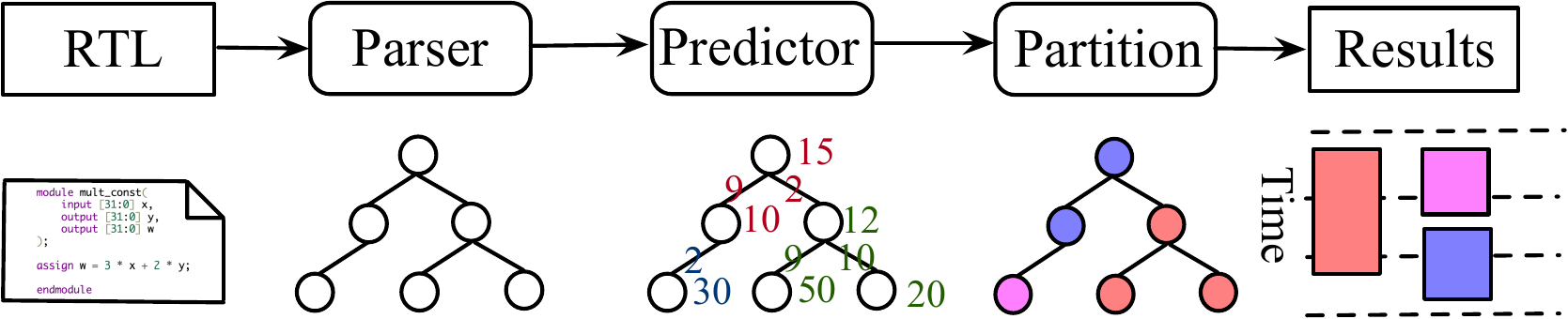} 
    \caption{Circuit Partition Pipeline}
    \label{fig:circuit-partition}
\end{figure}

An effective circuit partitioning approach should aim to achieve an optimal total synthesis time while minimizing the loss of performance. 
\Cref{fig:circuit-partition} shows the circuit partition pipeline.
The RTL code passes an initial transformation into an abstract syntax tree (ast) using a parser, then transformed to an instance tree, from which one node represents a Verilog module. 
Subsequently, a circuit predictor is employed to estimate the synthesis time for each node. 
Based on these predictions, the circuit partitioner is designed to divide the circuit in a manner that balances the total synthesis workload while minimizing the loss of performance, and a scheduling algorithm is employed to efficiently allocate synthesis time slots to the sub-circuits, ensuring optimal resource utilization and minimizing overall synthesis time.

\minisection{Circuit Predictor}
To evaluate the synthesis time of each module, we propose to establish a circuit predictor to predict the related synthesis time of each node. 
For RTL code, arithmetic and control operations such as     binary and unary operators, conditionals, and element-selects with varying bit-widths (e.g., Add, Equality, ArithmeticShiftLeft, ElementSelect) are heavily utilized for word-level and bit-level synthesis. After building and simplifying the Abstract Syntax Tree (ast), the tree is traversed, and the bit-widths of ast nodes are compiled into a high-dimensional feature vector for subsequent prediction.

We train our circuit predictor in an offline manner where real industry-level data is leveraged. We adopt XGBoost~\cite{chen2016xgboost} as the predictor and train the model to predict the node weights. 
Edge weight prediction involves estimating the performance, power, and area (PPA) effects using edge weights. These weights are calculated based on the type of connection between instances in a Verilog module, classified as direct, combinational, or sequential. This process supports workload prediction and chip partitioning.

\minisection{Circuit Partitioner}
To optimize parallel synthesis efficiency, we focus on partitioning the instance tree to balance synthesis time and PPA, which can be formulated as:
\begin{equation}
	\begin{aligned}
		\min_{S} \;\; & C = L + \lambda E \\
		s.t.\;\; & N_{min} \leq |S| \leq N_{max}, \\
	\end{aligned}
        \label{eq:circuit-partition}
\end{equation}
where $C$ is the total cost, $L$ represents overall synthesis time and $E$ denotes the cost of edgecut.
The edgecut is closely related to the granularity of the RTL code, and we generally prefer smaller granularity. 
$S$ is the set of edgecuts, where the cardinality of $S$ is equal to number of partitions.
$N_{min}$ and $N_{max}$ represent the max and min number of partitions.
By solving \Cref{eq:circuit-partition}, we aim to achieve balanced partitions with minimal synthesis time and optimal edgecut.

To address the problem, we propose a hierarchical tree partitioning algorithm. It starts by partitioning the instance tree in a top-down manner, where we partition the root node and associated sub-trees. Then, we evaluate the cost 
$C$ using a bin-packing algorithm with a first-fit strategy. 
Subsequently, We iteratively partition the sub-tree with the largest weight and evaluate the cost $C$.
The partitioning process continues until it meets the maximum partition number or the cost does not improve further.
After partition, we transform ast to Verilog Code for further rewriting.

\subsection{Large Models aided RTL Rewriting} \label{ssec:rtl-rewriting}
\begin{figure}[tb!]
    \centering    \includegraphics[width=1.\linewidth]{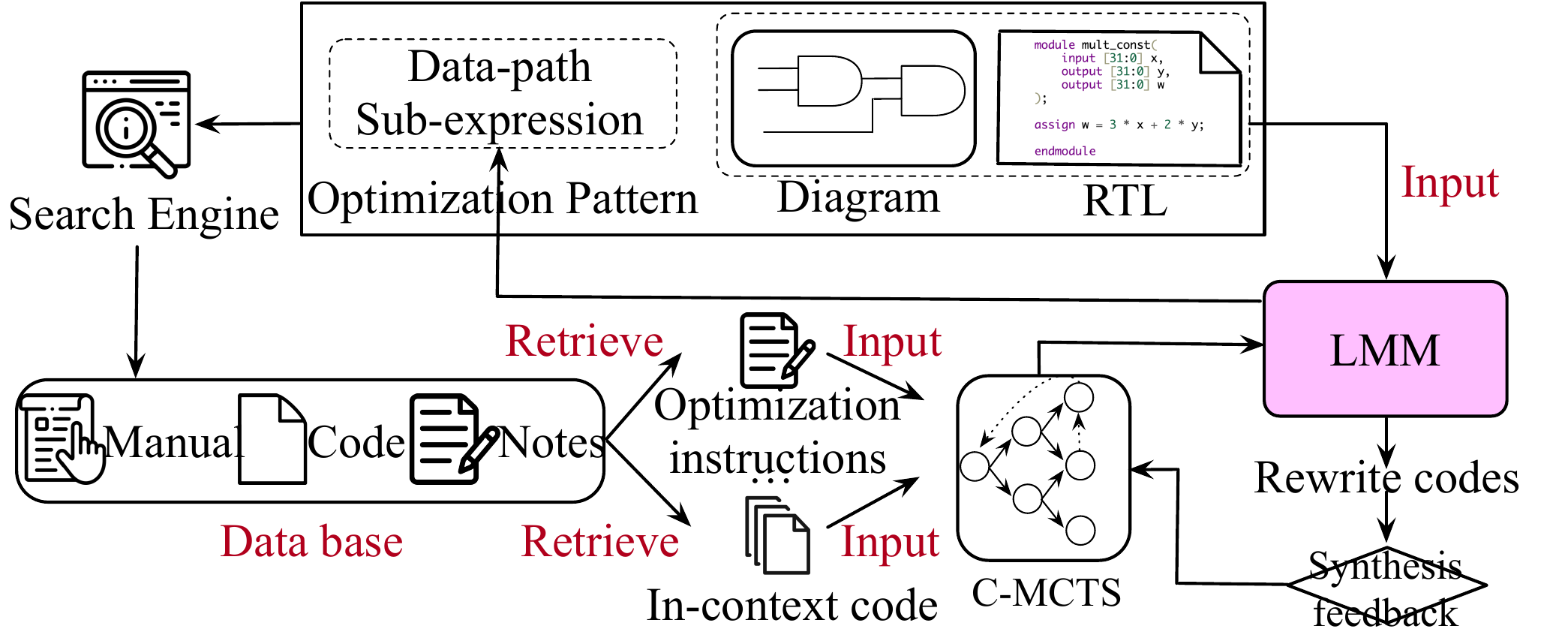} 
    \caption{RTL Rewriting Pipeline}
    \label{fig:rtl-rewrite}
\end{figure}

\Cref{fig:rtl-rewrite} illustrates the whole large models aided RTL rewriting pipeline. 
We first introduce a multi-modal program analysis pipeline to extract optimization and verification patterns.
Then the optimization pattern, along with RTL code and diagram are regarded as queries to a search engine. 
The retrieved contents from self-established database will act as optimization guides, thereby enhancing the capability of large models to generate improved rewriting code.
To better utilize various sources of retrieved contents, we also introduce the Cost-aware Monte Carlo Tree Seach algorithm to efficiently direct optimization path.

\begin{figure}[tb!]
    \centering    \includegraphics[width=1.02\linewidth]{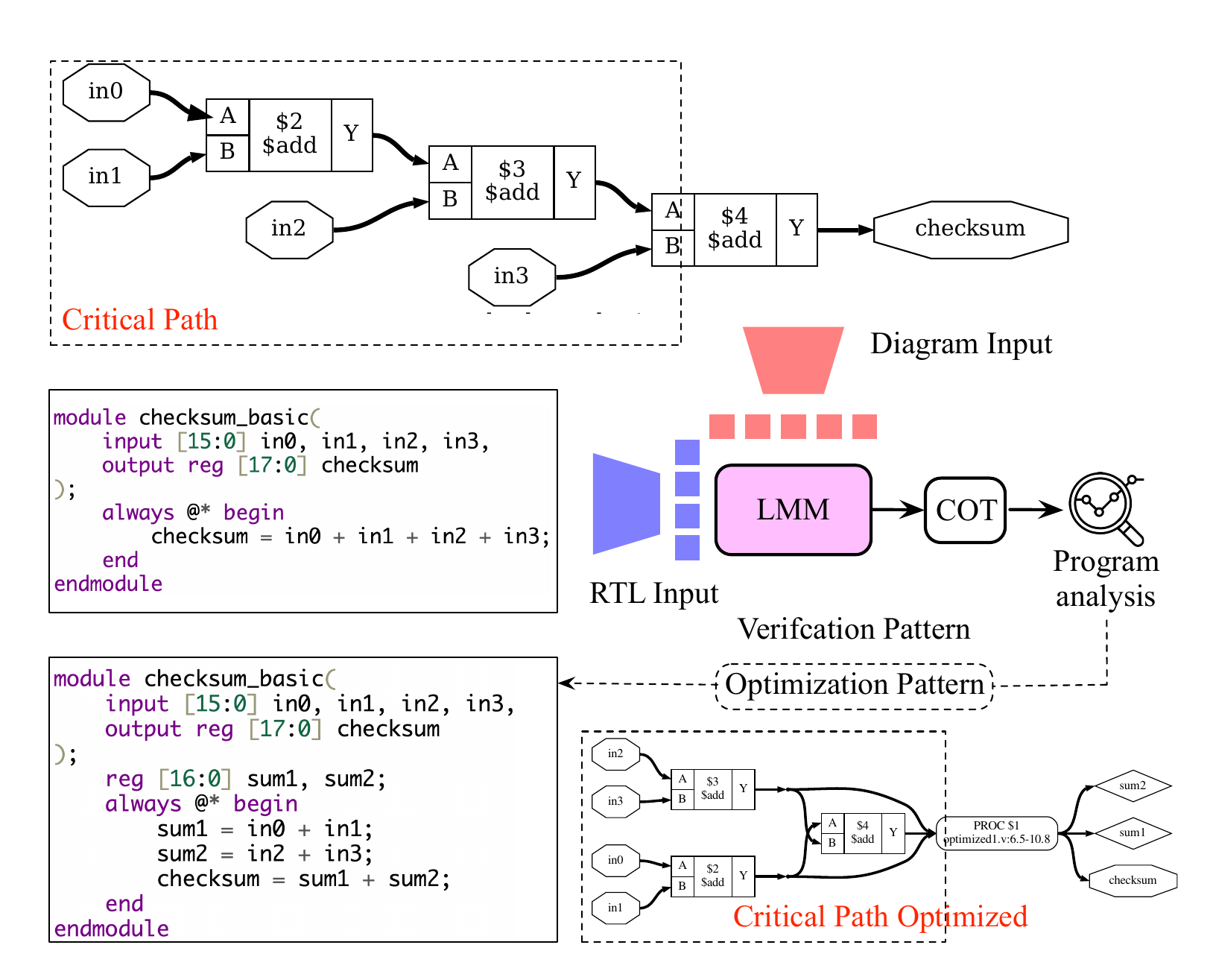} 
    \caption{Multi-modal Program Analysis }
    \label{fig:program-analysis}
\end{figure}

\minisection{Program Analysis and Pattern Recognition} \label{ssec:progam-analysis}
Program analysis serves as a crucial role in circuit design. 
However, current compiler-based approaches fall short in extracting semantic information from raw code.
For example, Visual information is an important clue in optimizing RTL code, ~\Cref{fig:program-analysis} shows a motivated visual example from which it's easy to recognize there existing optimization area for reducing critical path. 

Inspired by chain-of-thought (COT)~\cite{wei2022chain}, we propose a large multi-modal model (LMM) COT  program analysis pipeline. 
The pipeline first prompts large models to provide an in-depth analysis of diagrams based on the comprehension of given RTL code. 
The insights of proposed chain of analysis stem from the observation that large models can provide more accurate analysis of diagram when equipped with related RTL code, and combing diagram and RTL code can provide more informative and targeted guidance.
Defining large models as $\pi_{\theta}$, visual diagram as $x_v$, RTL code as $x_c$, initial prompt as $p_{init}$, and the optimization and verification patterns as $p_{opt}$ and $p_{ver}$, respectively, the process can be formulated as follows:

\begin{equation}
	\begin{aligned}
		p_{chain} ={}& \pi_{\theta}(x_v, x_c, p_{init}), \\
		p_{opt}, p_{ver} ={}& \pi_{\theta}(p_{chain}, x_v, x_c, p_{out}),
	\end{aligned}
\end{equation}
where $p_{chain}$ represents the chain of thoughts output, and $p_{output}$ denotes the final output prompt.

While program analysis can be applied to various scenarios in the RTL optimization process, this research primarily focuses on utilizing program analysis for optimization and verification pattern recognition. Given an RTL code and its associated diagram, we aim to leverage program analysis to identify relevant optimization patterns, such as data-path categorization and sub-expression elimination directions. Regarding verification, the large models aim to determine the type of circuit (e.g., combinational circuit, arithmetic) to enable the use of more appropriate solvers for fast verification, as detailed in~\Cref{ssec:fast-verification}.
The structured output is produced by the specifically designed prompt such as "please return in the following format <\texttt{\textasciigrave\textasciigrave\textasciigrave}Optimization pattern>your response<\texttt{\textasciigrave\textasciigrave\textasciigrave}>"  allowing for the extraction of key elements from specific patterns.

\minisection{Search Engine} \label{ssec:search_engine}
Retrieval augmented generation (RAG)~\cite{gao2023retrieval} effectively enhances the generation capabilities of large models. To facilitate this, we establish a RTL database that stores relevant diagrams, codes, optimization instructions, and algorithms. The details of database creation require a considerable amount of manual effort and are omitted due to page limits.

\Cref{table:X-RAG} presents the various retrieval types, their representations, and query methods employed in our approach. For diagrams, we store the embeddings generated by a large vision model. As visual inputs are not directly suitable for providing optimization instructions, we consider them as bridges connecting related optimization instructions and codes. Consequently, we employ a join query approach~\cite{ioannidis1996query}, where the visual input is first searched, and then its mapped optimization instructions and codes are retrieved.
For codes and optimization instructions, we primarily utilize the embeddings generated by language models. Additionally, we incorporate the traditional TF-IDF~\cite{aizawa2003information} representation for code search to enhance retrieval effectiveness. When it comes to algorithms, we adopt a join query method. In this approach, optimization instructions or codes are initially searched and then serve as a bridge to retrieve related algorithms.

For diagrams and optimization instructions, we use the widely adopted cosine distance as the similarity metric and retrieve the top-$k$ most associated instances. For codes, we design a two-stage ranking approach to identify the top-$k$ most relevant code instances for a given RTL code query.
In the first stage, for any query code $x_d$, we generate keyword and semantic vectors, $\mathbf{v}^w_i$ and $\mathbf{v}^s_i$, using the TF-IDF encoder and LLM, respectively. The similarity between $x_d$ and each instance $I_i \in D_c$ in the code database $D_c$ is defined as:
\begin{equation}\label{eq:sim_score}
    sim(x_d, I_i) = \lambda \cdot cosine(\mathbf{v}^w,\mathbf{v}^w_i) + (1-\lambda)\cdot cosine(\mathbf{v}^s,\mathbf{v}^s_i) +1,
\end{equation}
where $\lambda \in [0,1]$ is a hyper-parameter between semantic similarity and keyword similarity and $ sim(x_d, I_i) \in [0,2]$. 
We select the top-$N$ similar instances $\hat{D}^{x_d}_c$ from the code database $D_c$ based on the similarity score.
In the second-ranking stage, our goal is to pinpoint the top-$k$ relevant yet diverse code instances. These selections are aimed at providing a broader range of code patterns, which can help large models offer better optimizations.
Formally, given $N$ code instances $\hat{D}^{x_d}_c$ and the query RTL code $x_{d}$, we select the top-$k$ relevant and diverse code $D^{x_d}_c$ by maximizing the following objective: 
\begin{align}\label{eq:rerank_obj}
    \max_{D^{x_d}_c \subseteq \hat{D}^{x_d}_c}{\sum_{I_i \in D^{x_d}_c}{sim(x_{d}, I_i)} +   \frac{1}{k}\cdot \sum_{I_i \in D^{x_d}_c}{dis(I_i, D^{x_d}_c)} },
\end{align}
where distance $dis(I_i, D^{x_d}_c)=\min_{I_j \in D^{x_d}_c \setminus I_i}{(2-sim(I_i,I_j))}$ denotes the diversity value between each instance $I_i$ and the other instances in $D^{x_d}_c$ and $dis(I_i, D^{x_d}_c) \in [0,2]$.

\begin{table}[tb!]
    \caption{Different Retrieval types and Query methods}
    \resizebox{.9\linewidth}{!}
    {
        {
    \begin{tabular}{c|c|c}
    \toprule
    Type &  Representation & Query Method  \\ \midrule
    Diagram        & Visual encoder's embedding & Join Query \\
    Code       & TF-IDF \& LLM's embedding  & Direct Query                    \\
    Optimization Instruction        & LLM's embedding   & Direct Query                                    \\ 
    Optimization Algorithm       & Text string & Join Query             \\ \bottomrule
    \end{tabular}
    \label{table:X-RAG}
        }
    }
\end{table}

\minisection{C-MCTS}
Although search engines can retrieve a wealth of contents, including code, optimization instructions, and algorithms, we have observed that supplying large models with all the search results as context does not consistently yield satisfactory outcomes, and can sometimes even degrade performance.
~\Cref{table:RAG-content} shows different RAG context that proved to be effective in optimizing corresponding cases.
Specifically, certain contexts, such as algorithms, are crucial for optimizing scenarios like finite state machines (FSM), where state reduction heavily depends on the optimization analysis of state transition tables.

\begin{table}[tb!]
    \caption{Different RAG contents}
    \resizebox{1.02\linewidth}{!}
    {
        {
    \begin{tabular}{c|c|c}
    \toprule
     Case &  RAG content & Type  \\ \midrule
    Adder        & Illustrates the critical path & Diagram \\
    FSM       & State minimization algorithm description  & Algorithm                    \\
    Data-path        &  Sub-expression elimination   & Optimization hints                                         \\ 
    MUX       &  Similar Codes instances Codes & Code            \\ \bottomrule
    \end{tabular}
    \label{table:RAG-content}
        }
    }
\end{table}

\begin{figure}[tb!]
    \centering    \includegraphics[width=1.02\linewidth]{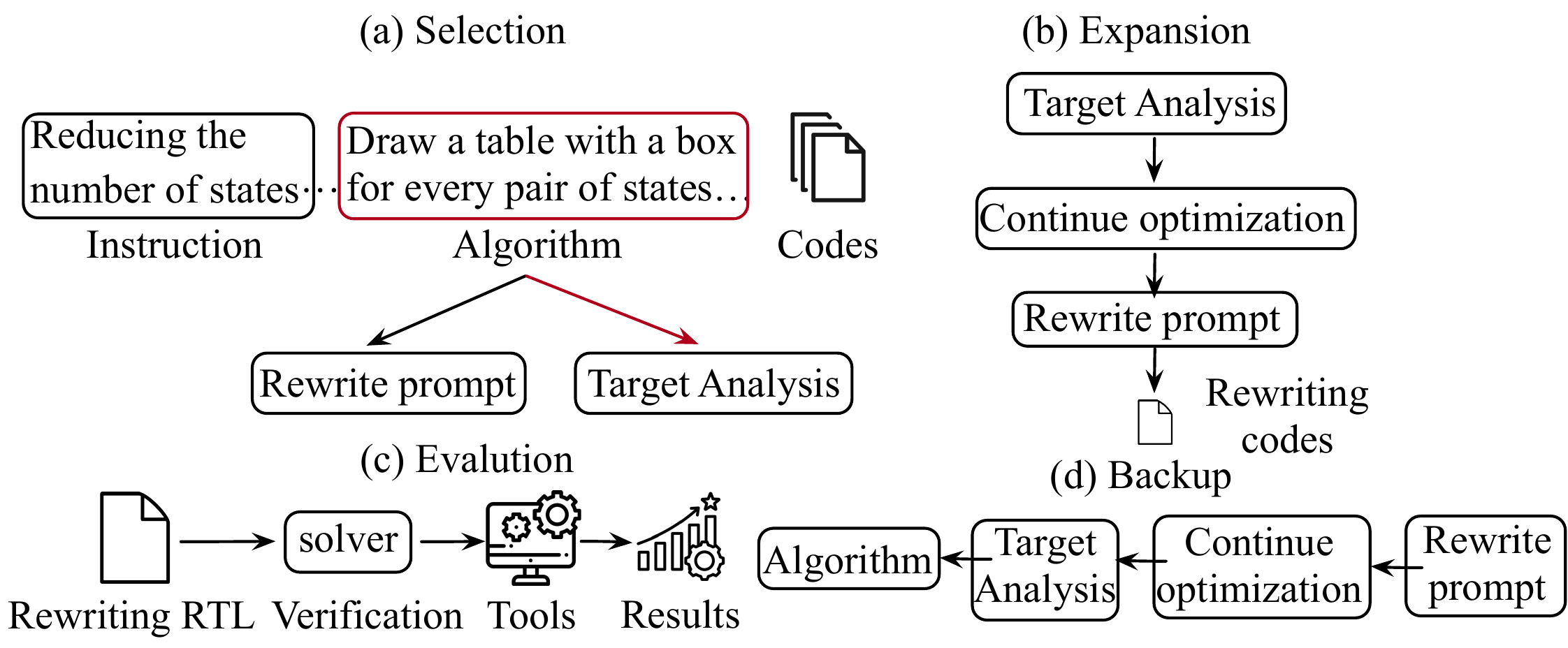} 
    \caption{C-MCTS Pipeline}
    \label{fig:method-mcts}
\end{figure}

Nevertheless, selecting the suitable retrieval content is not easy and sometimes the good generation results depend on combination of several retrieval contents. 
Additionally, an optimal rewriting instance often requires multiple iterations, which causes high inference cost of large models.
In our case, the initial selection of RAG contents involves 7 states, the rewriting codes are generated by several times (e.g., 10) due to the randomness of large models generation and the final rewriting code needs to be generated by multiple rounds in average. Consequently, the total number of states is over 1000 and it's impractical to utilize large models in a brute-force manner.
To address the problem, we propose a Cost-aware Monte Carlo Tree Search (C-MCTS) algorithm for proficient and efficient exploration while enabling strategic selection of retrieval contents and rewriting prompts.
The key elements of C-MCTS in our framework are illustrated as follows:

\begin{itemize}
[leftmargin=12pt]
    \item \textbf{State $s$}: represents current status of the problem. The initial state $s_0$ represents the initial selection of retrieval content combined with designed prompt and RTL code, while the intermediate states represent the all sending and generating contents in this state.
    \item \textbf{Action $a$}: represents the designed prompt that large models used for generation. We design four prompt templates in action as illustrated in~\Cref{table:action}.
    \item \textbf{Value function $Q(s, a)$}: denotes values of state-action pairs $(s, a)$ based on the synthesis results of rewriting RTL code. 
    \item \textbf{Cost function $C(s, a)$}: estimates large models inference cost in state-pair $(s,a)$.
\end{itemize}
As shown in~\Cref{fig:method-mcts}, the C-MCTS algorithm encompasses four key phases: selection, expansion, evaluation and propagation.
We provide details of each process below:

\minisection{Selection}:
In the selection phase, the algorithm begins at the root state and proceeds to choose an action $a^{*}$ from designed action set $\mathcal{A}$ as shown in~\Cref{table:action}.
The initial root state contains 7 branches involving all combinations of retrieved content. To encourage lower-cost inference, we assign a higher initial value to the states with less retrieved content. For instance, the state value is set to $1$ for retrieved code instances only, $\frac{1}{2}$ for the combination of optimization instructions and code instances, and $\frac{1}{3}$ for the combination of optimization instructions, code instances, and algorithms.
The selection aims to choose the most appropriate state to continue rewriting with minimal cost estimation, as follows:
\begin{equation}
a^* = \argmax_{a} (Q(s,a) + \lambda U(s,a) + \gamma C(s,a)),
\end{equation} 
where $Q(s,a)$ denotes the expected reward for the current state, $U(s,a)$ quantifies the associated uncertainty~\cite{auer2002using}, $C(s,a)$ represents the estimated cost function.

For each state, the value is based on its rewriting results. 
if the rewriting code can pass the verification and obtain better synthesis results then original code, the value is 1, otherwise the reward is 0.5.
The value is set to 0 if the rewriting code does not pass the verification, given by:
\begin{equation}
	Q(s,a) = 
	\begin{cases} 
    0 & \text{verification failed}, \\
    0.5 & \text{equivalent results lower than original RTL}, \\
    1 & \text{equivalent results surpass original RTL},
   \end{cases}
   \label{eq:dynamic-score}
\end{equation}
The value of state $s$ is then updated by summation of previous values and then divided by number of selections of this state.

\begin{table}[tb!]
    \caption{Actions Description}
    \resizebox{1.02\linewidth}{!}
    {
        {
    \begin{tabular}{c|c}
    \toprule
    Action &  Description   \\ \midrule
    Rewrite prompt      & Directly rewriting RTL code. \\
    Optimization analysis       & Analyze  optimization target, e.g., state reduction.                    \\
    Continue optimization        &  Prompt large models to continue optimizing targets.                                         \\ 
    Reflection       &  Carefully self-checking the rewriting codes.            \\ \bottomrule
    \end{tabular}
    \label{table:action}
        }
    }
\end{table}

Uncertainty function $U$ is given by:
\begin{equation}
	U(s,a) = \sqrt{\frac{2\log (T)}{N(s,a)}}, 
  \label{eq:ucb}
\end{equation}
where $T$ is total selection times, $N(s,a)$ is the number of selections of this state branch. 
~\Cref{eq:ucb} is the classical UCB algorithm that can be derived by Hoeffding's inequality~\cite{auer2002using}.
As $N(s,a)$ increases, the uncertainty of related state decreases.
Uncertainty measures the exploration degree, it the associated factor $\lambda$ is larger, we encourage more exploration and vice versa.

For the cost function, we consider both the historical cost and the estimated cost for successful rewriting in the future, defined as follows:
\begin{equation}
    \begin{aligned}
        C(s,a) ={}&  v(s,a) \cdot (1 + w(s,a)), \;\; \text{and} \\
        w(s,a) ={}& \frac{1}{{1 + e^{-10 \cdot (Q(s,a) - k)}}}.
    \end{aligned}
\end{equation}
where $v(s, a)$ denotes the success rate of state $s$. Denote the number of rewriting results that can pass the verification as $n$, then the success rate is calculated by $\frac{n}{N(s, a)}$. $w(s, a)$ measures the estimated cost to successfully rewrite the code.
$w(s, a)$ is designed like a reinforced sigmoid function and $k$ is the balance score factor.
If the value of this state $Q(s,a)$ exceeds the balance score factor $k$, we anticipate a lower cost for successful rewriting, and thus lower the estimated cost accordingly. Conversely, if the value of the state
$Q(s)$ is less than $k$, the estimated cost is increased.

\minisection{Expansion and Evaluation}:
After the selection phase, we expand to the state $s$ for further exploration. Instead of training a policy network like in traditional reinforcement learning, we leverage large models to decide the next action based on pre-designed options, as shown in \Cref{table:action}. The action selection depends on the structured output generated by the large models for state $s$ using a specially designed prompt, similar to those used in program analysis. The expansion continues until a rewriting action is chosen. After rewriting, we use synthesis tools to evaluate performance and check equivalence between the original and rewritten code.

\minisection{Backpropagation}: After obtaining verification and synthesis results, we update the value of state $s$ as illustrated in~\Cref{eq:dynamic-score}.

We iteratively adopt each process and append the rewriting codes that can pass verification and surpass original Code's synthesis results in a list. 
When there's no obvious improvement or meeting defined selection times, we stop the search process and choose the best rewriting instances.

\subsection{Fast Verification} \label{ssec:fast-verification}

Verification is a crucial part in circuit design, ensuring the correctness of the implemented functionality.
In our framework, verification becomes even more crucial due to the inherent randomness generation by large models.
In the realm of RTL optimization frameworks, a primary focus lies in equivalence checking~\cite{lavagno2016electronic}. This process aims to validate the functional equivalence between the original Verilog code and its rewritten version.
Modern open-source compilers mainly use SAT-based approaches~\cite{marques1999combinational}, which translate the equivalence problem into a SAT problem. 
If the SAT finds the problem unsatisfiable, the RTL codes are not equivalent.
However, the worst-case computational complexity of SAT is NP-complete, which means the time required to solve a problem can grow exponentially with the size of problem.
For example, as shown in~\Cref{fig:fast-verfication}, this arithmetic circuits lead to unexpected verification times due to the complexity of the Boolean representations of such operations.
In contrast, this kind of arithmetic circuit can be efficiently solved by other approaches such as symbolic algebraic methods~\cite{sayed2016equivalence}. 
These approaches leverage algebraic techniques and tools to simplify and compare these arithmetic expressions. 

\begin{figure}[tb!]
    \centering    \includegraphics[width=1.\linewidth]{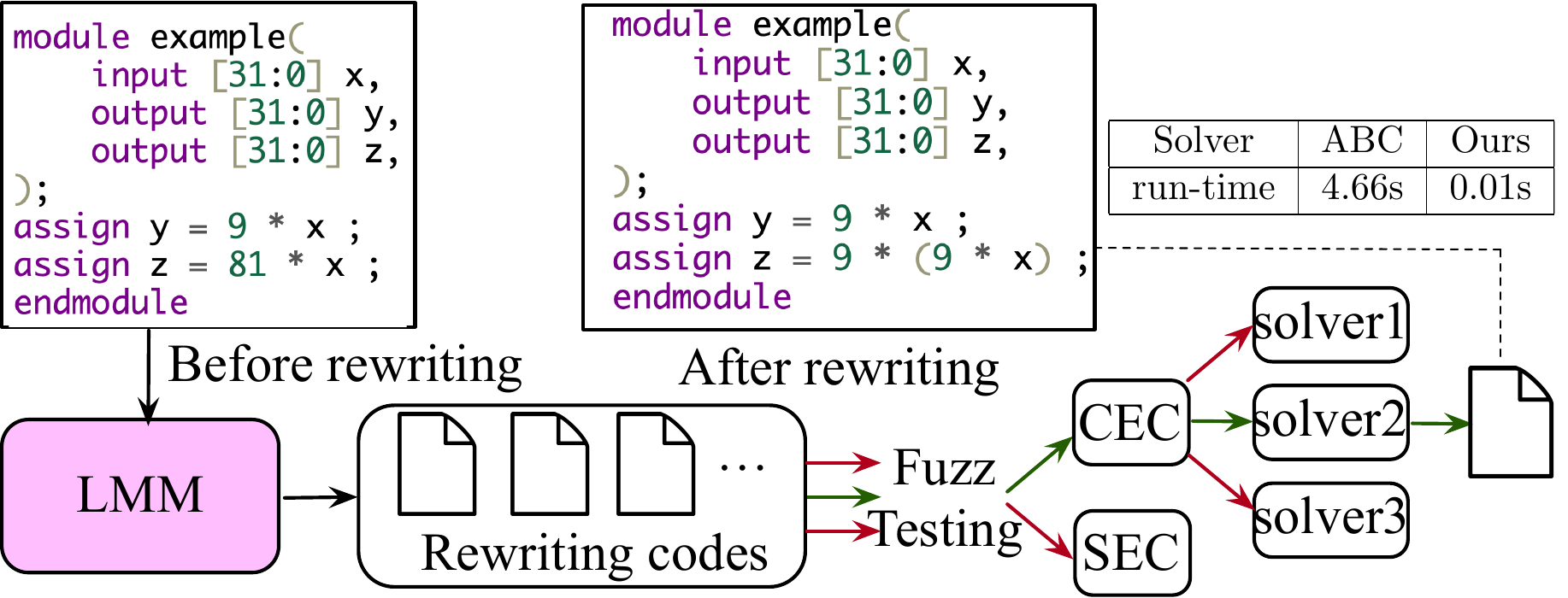} 
    \caption{Fast Verification Pipeline}
    \label{fig:fast-verfication2}
\end{figure}

Due to the randomness of large model-generated rewritten codes, conducting equivalence checking for all cases is costly. To address this, we leverage fuzz testing~\cite{godefroid2008automated}, which generates random test cases to simulate Verilog codes. By comparing the outputs of the original and rewritten code, we can identify differences and filter out problematic cases, reducing the need for extensive verification. This approach helps optimize the RTL optimization framework by avoiding the burden of exhaustive verification.

\Cref{fig:fast-verfication2} illustrates our fast verification pipeline. After generating multiple rewritten codes via large models, we employ open-source tools to randomly generate test cases. We then filter out those rewritten cases whose outputs differ from the original. Subsequently, we utilize the verification pattern generated by program analysis as detailed in~\Cref{fig:program-analysis} to determine the Verilog code is combinational or sequential and choose the appropriate solver.


\section{Experiments}

\subsection{Experimental Settings} \label{ssec:exp-settng}

\minisection{Benchmarks}
In this study, we introduce two benchmarks for RTL code optimization, designed to address both long and short code scenarios, crafted by experienced Verilog engineers. These benchmarks derive their optimization patterns from a comprehensive review of internal industry documents and a survey of approximately 50 scholarly articles on RTL optimization.

The benchmarks encompass real-world scenarios, refined to illustrate typical industry challenges, with optimized solutions provided by skilled Verilog engineers. For the short benchmark suite, out of an initial 55 cases, 14 representative scenarios were selected for detailed analysis in our experimental results due to page constraints. These scenarios cover various RTL code aspects, including basic patterns, data-path, memory, MUX, FSM, and control logic.
The long benchmarks include five extensive cases involving CPU design, neural networks, and image processing, which is more challengeable.


To facilitate bridging the gap between academic research and industrial practice in RTL code optimization, the benchmarks are  at \href{https://github.com/yaoxufeng/RTLRewriter-Bench}{https://github.com/yaoxufeng/RTLRewriter-Bench}.

\minisection{Baselines}
We compare RTLRewriter with 6 baselines in three types approaches as follows:
\begin{itemize}
[leftmargin=12pt]
    \item \textbf{Two SOTA large models}: We evaluate RTLRewriter in comparison to two leading large-scale models including GPT4~\cite{achiam2023gpt} and Claude3-Opus. We access GPT-4 and Claude3-Opus via Poe (https://poe.com).  
    \item \textbf{Two SOTA open-source large RTL models}: RTLRewriter is also compared with two specialized language models designed for RTL code generation:
    VeriGen~\cite{thakur2023verigen} and RTLCoder~\cite{liu2023rtlcoder} are two currently sota large language models targeting RTL code generation. 
    \item \textbf{Two competitive compilers}: Additionally, we assess RTLRewriter against two widely-used open-source compilers: Egg~\cite{coward2022automatic} and Yosys~\cite{wolf2013yosys}.
\end{itemize}
To ensure fair and consistent evaluation, all language model baselines are tested using \textbf{identical prompt engineering techniques}. For the compilers, we utilize the standard optimization commands in Yosys. We implement our own version of Egg integrated with Yosys, incorporating custom optimization optimization operators, which serves as a stronger baseline than Egg alone.

\begin{table*}[tb!]
    \centering
    \footnotesize
    \caption{Comparisons of baseline approaches on short RTL rewriting benchmarks}
    \label{tab:short-bench}
    {
        {
    \begin{tabular}{c|cc|cc|cc|cc|cc|cc|cc}
        \toprule
         Test Case & \multicolumn{2}{c|}{Yosys} & \multicolumn{2}{c|}{Yosys+Egg} & \multicolumn{2}{c|}{GPT4}& \multicolumn{2}{c|}{Claude3} & \multicolumn{2}{c|}{VeriGen} & \multicolumn{2}{c|}{RTLCoder}& \multicolumn{2}{c}{RTLRwriter} \\
         ID&Wires  & Cells &Wires & Cells &Wires  & Cells &Wires  & Cells&Wires  & Cells & Wires  & Cells & Wires  & Cells \\
        \midrule
         case1  & 28 & 18 & 24& 14&\textcolor{gray6}{28} &\textcolor{gray6}{18} & \textcolor{gray6}{28} & \textcolor{gray6}{18} & \textcolor{gray6}{28}& \textcolor{gray6}{18}&\textcolor{gray6}{28} & \textcolor{gray6}{18}& 24 & 14 \\
         case2  & 11646 & 11824 & 11307 & 11478  & 11507& \textcolor{gray6}{11684}& 11609 & 11787 &\textcolor{gray6}{11684} & \textcolor{gray6}{11824}&11588 & 11766& 11299 & 11477 \\
         case3  & 1136 & 1220 & 1095 & 1176 & 890 & 974 & 910 & 992  &\textcolor{gray6}{1135} &\textcolor{gray6}{1120} & \textcolor{gray6}{1136}&\textcolor{gray6}{1120} & 890 & 974 \\
         case4  & 1376 & 1462 & 1327 & 1409 &\textcolor{gray6}{1376} & \textcolor{gray6}{1462}& \textcolor{gray6}{1376} &\textcolor{gray6}{1462}  & \textcolor{gray6}{1376}&\textcolor{gray6}{1462} & \textcolor{gray6}{1376}& \textcolor{gray6}{1462}& 1127 & 1213  \\
         case5  & 193 & 49 & 164 & 49 &\textcolor{gray6}{193} & \textcolor{gray6}{49} & \textcolor{gray6}{193}  & \textcolor{gray6}{49} & \textcolor{gray6}{193}& \textcolor{gray6}{49}& \textcolor{gray6}{193}& \textcolor{gray6}{49}& 65 &\textcolor{gray6}{49} \\
         case6  & 172 & 129 & 161 & \textcolor{gray6}{129}& 171&\textcolor{gray6}{129} & 171 & \textcolor{gray6}{129} &\textcolor{gray6}{172} &\textcolor{gray6}{129} &\textcolor{gray6}{172} & \textcolor{gray6}{129}& 161 & \textcolor{gray6}{129}\\
         case7  & 402  & 403 & 378 & 379 & \textcolor{gray6}{402}& \textcolor{gray6}{403}&  386& 387& \textcolor{gray6}{402} & \textcolor{gray6}{403} &\textcolor{gray6}{402} & \textcolor{gray6}{403} &353 &354 \\
         case8  & 466 & 354 & 434 & 330 &\textcolor{gray6}{466} &\textcolor{gray6}{354} & \textcolor{gray6}{466} & \textcolor{gray6}{354} & \textcolor{gray6}{466}& \textcolor{gray6}{354}& \textcolor{gray6}{466}&\textcolor{gray6}{354} &370 & \textcolor{gray6}{354}\\
         case9  & \textcolor{gray6}{70} & \textcolor{gray6}{71} & 67 & 68 & \textcolor{gray6}{70} & \textcolor{gray6}{71} & \textcolor{gray6}{70} &\textcolor{gray6}{71}& \textcolor{gray6}{70} &\textcolor{gray6}{71} &\textcolor{gray6}{70} &\textcolor{gray6}{71} &34 & 32\\
         case10  & 59 & 56 &56 & 54 & \textcolor{gray6}{59}& \textcolor{gray6}{56}&  \textcolor{gray6}{59}& \textcolor{gray6}{56} &\textcolor{gray6}{59} &\textcolor{gray6}{56} &\textcolor{gray6}{59} &\textcolor{gray6}{56} & 41&42 \\
         case11  & 34 & 35 & 33& 34& 21&24 & \textcolor{gray6}{34} & \textcolor{gray6}{35} & \textcolor{gray6}{34}& \textcolor{gray6}{35}& \textcolor{gray6}{34}& \textcolor{gray6}{35}&21 &24 \\
         case12 & 14782 & 14960 & \textcolor{gray6}{14782}& \textcolor{gray6}{14960}& \textcolor{gray6}{14782} & \textcolor{gray6}{14960}& 14695 & 14873 &\textcolor{gray6}{14782} &\textcolor{gray6}{14960} &\textcolor{gray6}{14782} & \textcolor{gray6}{14960}&14525  & 14703 \\
         case13  & 7 & 2 &\textcolor{gray6}{7} &\textcolor{gray6}{2} &3 & 1& 3 &  1& \textcolor{gray6}{7}& \textcolor{gray6}{2} & \textcolor{gray6}{7} &\textcolor{gray6}{2} &3 &1 \\
         case14  & 16 & 6 &14 &5 &8 & 3& 8 & 3 & \textcolor{gray6}{16}& \textcolor{gray6}{6}& \textcolor{gray6}{16}& \textcolor{gray6}{6}&8 &3 \\
         \midrule
         GeoMean  & 222.68  & 161.97 & 209.14 & 153.22 & 189.19 & 140.43 & 195.56 & 144.04 & 222.68& 161.97& 222.60& 161.91&152.83 &124.46 \\
         Ratio  & 1.00 & 1.00 & 0.93 & 0.94 & 0.85 & 0.87& 0.88 & 0.89  & 1.00 &1.00 &0.99 &0.99 &  0.69& 0.77  \\
        \bottomrule
    \end{tabular}
        }
    }
\end{table*}

\begin{table*}[tb!]
    \centering
    \footnotesize
    \caption{Comparisons of baseline approaches on long RTL rewriting benchmarks}
    \label{tab:long-bench}
    \renewcommand{\arraystretch}{1.08}
    \resizebox{1.0\linewidth}{!}
    {
        {
    \begin{tabular}{c|cc|cc|cc|cc|cc|cc|cc}
        \toprule
         Test Case & \multicolumn{2}{c|}{Yosys} & \multicolumn{2}{c|}{Yosys+Egg} & \multicolumn{2}{c|}{GPT4}& \multicolumn{2}{c|}{Claude3} & \multicolumn{2}{c|}{VeriGen} & \multicolumn{2}{c|}{RTLCoder}& \multicolumn{2}{c}{RTLRwriter} \\
         ID&Area  & Delay &Area & Delay &Area  & Delay &Area  & Delay&Area  & Delay & Area  & Delay & Area  & Delay  \\
        \midrule
         CPU  & 179025.72 & 1989.76 & 167996.88 & 1688.54 &\textcolor{gray6}{179025.72} & \textcolor{gray6}{1989.76}& \textcolor{gray6}{179025.72} & \textcolor{gray6}{1989.76} & \textcolor{gray6}{179025.72}&\textcolor{gray6}{1989.76} & \textcolor{gray6}{179025.72} &  \textcolor{gray6}{1989.76} & 167634.27 & 1592.58\\
         CNN  & 26071.46 & 15890.42 & 22004.64 & 14746.34 & \textcolor{gray6}{20104.01}& \textcolor{gray6}{15890.42} & \textcolor{gray6}{20104.01} & \textcolor{gray6}{15890.42} & \textcolor{gray6}{20104.01} & \textcolor{gray6}{15890.42} & \textcolor{gray6}{20104.01} & \textcolor{gray6}{15890.42} & 20104.01 & 13565.95  \\
         FFT  & 71385.35 & 184098.72 & 60321.54 & 183545.68 &\textcolor{gray6}{71385.35} &\textcolor{gray6}{184098.72} & \textcolor{gray6}{71385.35} &\textcolor{gray6}{184098.72}  & \textcolor{gray6}{71385.35} & \textcolor{gray6}{184098.72} &\textcolor{gray6}{71385.35} &\textcolor{gray6}{184098.72} & 56451.58 & 181495.83 \\
         Huffman  & 106045.69 & 1544.00 & 97480.22 & 1544.36 & \textcolor{gray6}{106045.69} &\textcolor{gray6}{1544.00} & \textcolor{gray6}{106045.69} & \textcolor{gray6}{1544.00} & \textcolor{gray6}{106045.69}& \textcolor{gray6}{1544.00}& \textcolor{gray6}{106045.69}& \textcolor{gray6}{1544.00}& 99142.98 & 1545.64\\
         VMachine & 1212.43 & 569.20 & 1030.23 & 642.87 & \textcolor{gray6}{1212.43}& \textcolor{gray6}{569.20}&  \textcolor{gray6}{1212.43} &\textcolor{gray6}{569.20}  &  \textcolor{gray6}{1212.43} & \textcolor{gray6}{569.20} &  \textcolor{gray6}{1212.43} & \textcolor{gray6}{569.20} &799.60  &676.81  \\
         \midrule
         GeoMean  & 33602.22 &  5517.98 & 29513.98 & 5387.19 &33602.22&  5517.98& 33602.22 & 5517.98  & 33602.22&  5517.98&  33602.22 & 5517.98 & 27270.39 & 5279.57 \\
         Ratio  &1.00  & 1.00 & 0.87 & 0.97 & 1.00& 1.00& 1.00 &1.00  & 1.00 & 1.00 & 1.00 & 1.00 & 0.81 & 0.96  \\
        \bottomrule
    \end{tabular}	
        }
    }
\end{table*}

\minisection{Implementations}
In this research, we introduce the RTLRewriter framework, which leverages GPT-4V~\cite{achiam2023gpt}, a state-of-the-art multi-modal and language model, for multi-modal program analysis and RTL code optimization. We access GPT-4V via Poe using the following link: https://poe.com/GPT-4-Turbo. For circuit partitioning, we utilize Vivado to evaluate the PPA of the designs. Our search engine components include ViT~\cite{dosovitskiy2020image} as the image encoder, LLama3~\cite{touvron2023llama} for text embeddings, and DeepSeek~\cite{bi2024deepseek} for code embeddings. For verification, we employ iVerilog~\cite{williams2002icarus} to compile the code, alongside self-developed scripts for generating test cases. Additionally, we utilize ABC~\cite{brayton2010abc} and egg as our solvers, where ABC excels in SAT-based verification, while egg is highly efficient for arithmetic circuit verification.

\minisection{Evaluation Metrics}
For the overall optimization system, we focus on synthesis results and whole run-time. We also take other metrics that measures the effectiveness of each part. The calculations of these metrics are listed below:
\begin{itemize}
    [leftmargin=12pt]
    \item \textbf{Wires}: represent the interconnections between components, with a higher count indicating more complex routing.
    \item \textbf{Cells}: represent the logical components used in the design, with a higher count indicating more logic complexity.
    \item \textbf{Area}: total hardware resources utilized by the synthesized circuit. This includes the number of logic gates, flip-flops, and interconnections required to implement the design, generated by ABC~\cite{brayton2010abc}.
    \item \textbf{Delay}: measures the longest propagation time of signals through a circuit from input to output, generated by ABC~\cite{brayton2010abc}.
    \item \textbf{H@K}: Hit ratio for top-k retrieval results on $n_t$ RTL code queries.
    $H@K=\frac{1}{n_t}\sum_{i=1}^{n_t}\frac{1}{K}\sum_{k=1}^{K}\mathbb{I}(y_{i,k},y_{i})$,
    where $y_{i,k}$ denotes the code type of the retrieved $k$-th code for query code $x_{di}$, and the indicator function $\mathbb{I}(y_{i,k},y_{i})=1$ if $y_{i,k}=y_{i}$.
    \item \textbf{MAP@K}: Mean average precision (MAP) for top-k results on $n_t$ RTL code queries. 
    $MAP@K=\frac{1}{n_t}\sum_{i=1}^{n_t}\frac{1}{K}\sum_{k=1}^{K}\frac{\mathbb{I}(y_{i,k},y_{i})\cdot n(x_{di,\le k})}{k}$,
    where $n(x_{di,\le k})$ denotes the number of code instances in the first top-$k$ that has the same code category with query $x_{di}$.
    
\end{itemize}

\subsection{Performance Analysis} \label{ssec:performance}
\Cref{tab:short-bench} illustrates the results of short RTL rewriting benchmarks across 16 cases, encompassing both baseline results and human implementations, validated by senior Verilog engineers. Wires and cells are two critical evaluation metrics in front-end RTL optimization due to their strong correlation with area and delay. In real-world applications, optimizing these metrics, even by 1\%, represents significant progress.

\begin{figure}[tb!]
    \centering    
    \includegraphics[width=.86\linewidth]{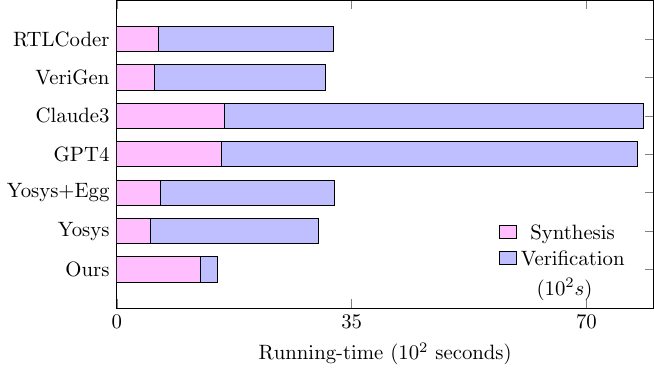} 
    \caption{Running-time Comparison}
    \label{fig:fast-verfication}
\end{figure}

We adopt Yosys results as our baseline, where fewer wires and cells indicate better results. 
The results with gray color mean that the method fails to optimize the code.
Notably, our method outperforms all competitive baselines by a substantial margin. Among these cases, state-of-the-art large models like GPT4 and Claude3-Opus exhibit great potential in optimizing RTL code, even surpassing curated compiler-based solutions such as Yosys+Egg. However, without the aid of additional techniques, these raw large models struggle to optimize diverse cases. 
Additionally, we have observed that the large RTL models targeting code generation such as~\cite{liu2023rtlcoder} and VeriGen~\cite{thakur2023verigen} can hardly optimize the RTL code even they are fine-tuned from large corpus of RTL dataset.
In contrast, our method achieves optimization in nearly all cases.
Interestingly, in some cases, such as case 14, the large models generate a superior implementation compared to the human version, underscoring the potential for leveraging large models in RTL code optimization.

\Cref{tab:long-bench} presents the results for the long-benchmark. These cases are designed to simulate real industry applications, featuring significantly longer code lines. The evaluation metrics are synthesis area and delay obtained via vivado.
We observe that the proposed RTLRewriter continues to outperform all baseline methods by a substantial margin. Due to the complexity of the designs, GPT-4 and Claude 3-Opus were less effective in optimizing RTL code. However, the Yosys+Egg approach consistently demonstrated improvements.
Unlike the short cases, the long cases are more complex, resulting in smaller overall improvements. Nevertheless, these results highlight the potential for applying large models, coupled with dedicated strategies, to optimize RTL code.

 Another important evaluation metric in RTL optimization scenarios is running time, which encompasses both synthesis time and verification time. As shown in \Cref{fig:fast-verfication}, our method achieves significantly faster overall running time. Compared with other approaches based on large models such as GPT4 and Claud3-Opus, our method achieves much faster synthesis time, thanks to our fuzz testing strategy. Compared to other compiler-based methods, our approach significantly reduces verification time, especially for arithmetic circuits where the large UNSAT core can consume considerable time.
Large RTL models like VeriGen and RTLCoder achieve similar synthesis and verification runtimes as compiler-based methods, largely because they often generate repetitive original RTL code. Despite this, our method consistently outperforms these approaches.

\subsection{Ablation Studies} \label{sec:ab-study}
In this subsection we provide more ablation studies on each part within our framework.
\Cref{table:circuit-partition} presents the ablation studies on circuit partition.
We observe that without partitioning, the synthesis time is ten times longer than with partitioning, owing to the parallel synthesis strategy. Additionally, the PPA loss is almost negligible compared to the w/o partition approach. More importantly, when testing on the long-benchmark, we found that it is nearly impossible for large models such as GPT4 or Claude3-Opus to optimize the entire code correctly, highlighting the importance and effectiveness of the partitioning strategy.

\begin{table}[tb!]
    \centering
    \footnotesize
    \caption{Ablation on Circuit Partition}
    {
        {
	    \begin{tabular}{c| c | c|c}
	    \toprule
	     Method &  Synthesis-time & PPA & Rewriting Performance\\ \midrule
	    w/o Partition         & 10.12  & 1.00 & 1.00\\ 
	    Ours        &  1.00 & 0.99 & 0.82\\ \bottomrule
	    \end{tabular}
	    \label{table:circuit-partition}
        }
    }
\end{table}

Multi-modal program analysis is a crucial component of our framework, where integrating other modalities, such as visual information, is naturally aligned with RTL code analysis. \Cref{tab:rewrite-finding,tab:verification-finding} present a comparison between GPT-4 and our method, on small benchmark.
It is evident that our method achieves better pattern-finding performance than GPT-4 on both RTL optimization pattern finding and verification pattern finding.

\begin{table}[tb!]
    \centering
    \begin{minipage}{0.45\linewidth}
        \centering
        \footnotesize
        \caption{RTL Optimization Pattern Finding}
        {
            \begin{tabular}{c| c | c}
            \toprule
             Method &  Precision & Recall \\ \midrule
            GPT4         & 64.12\%  & 72.15\% \\ 
            Ours        & 86.14\%  & 83.12\% \\ \bottomrule
            \end{tabular}
        }
        \label{tab:rewrite-finding}
    \end{minipage}
    \hspace{0.05\linewidth}
    \begin{minipage}{0.45\linewidth}
        \centering
        \footnotesize
        \caption{Verification Pattern Finding}
        {
            \begin{tabular}{c| c | c}
            \toprule
             Method &  Precision & Recall \\ \midrule
            GPT4         & 96.64\%  & 92.23\% \\ 
            Ours        & 99.16\%  & 97.24\% \\ \bottomrule
            \end{tabular}
        }
        \label{tab:verification-finding}
    \end{minipage}
\end{table}

~\Cref{tab:search_engine} shows the effectiveness of our search method compared to other baselines, including traditional methods such as TF-IDF~\cite{aizawa2003information}, BM25~\cite{robertson2009probabilistic}, and machine learning methods such as XGBoost~\cite{chen2016xgboost} and RandomForest~\cite{liaw2002classification}.
Our method surpasses all baselines in accurately retrieving code instances, demonstrating its superior ability to decode complex RTL code patterns beyond the capabilities of traditional and machine learning models. Traditional methods like TF-IDF and BM25 lack the depth to understand complex code patterns, while models like XGBoost and RandomForest fall short in semantic comprehension. Our engine effectively combines textual and semantic analysis, enhancing retrieval performance.

\begin{table}[tb!]
    \centering
    \footnotesize
    \caption{Evaluation on search Engine
    }
    {
        \begin{tabular}{c|cc|cc}
            \hline
            & H@1 & H@5  & MAP@3 &MAP@5  \\ \hline
            \textit{Optimal} & \textit{1.00}      &  \textit{1.00}    & \textit{1.00} &  \textit{1.00}        \\ \hline

            TF-IDF & 0.88       &  0.46    & 0.46 &  0.61       \\ 
            BM25 & 0.82   & 0.48    &   0.41   & 0.72    \\  
            XGB &    0.98 &  0.31   &  0.26    & 0.44        \\ 
            RF &   0.72  &  0.28   &  0.22   &   0.34     \\ \hline
            Ours &  \textbf{1.00}    &  \textbf{0.68}    &  \textbf{0.76}  & \textbf{0.88}       \\ \hline
        \end{tabular}
        \label{tab:search_engine}
    }
\end{table}

\Cref{table:c-mcts} compares our proposed C-MCTS with other baselines. We used GPT-4 for all baselines with specific selection times for a fair comparison.
We mainly test our results on small benchmark.
GPT-4-DFS uses prompt engineering with multiple iterations, which improves its performance over standard GPT-4. However, GPT-4-MCTS and our C-MCTS enable more effective exploration, resulting in greater improvements.
Overall, our method surpasses all baselines by a significant margin, demonstrating the effectiveness of our approach.

\begin{table}[tb!]
    \centering
    \footnotesize
    \caption{C-MCTS Comparison}
    \label{table:c-mcts}
    {
    \begin{tabular}{c|c c c c}
    \toprule
     Method &  GPT4 & GPT4-DFS & GPT4-MCTS & Ours \\ \midrule
    PPA-ratio   & 1.0  & 0.93 & 0.84 & 0.81\\  \bottomrule
    \end{tabular}
    }
\end{table}

\section{Conclusion}

Traditional RTL code optimization relies heavily on the expertise of skilled engineers, often requiring multiple iterations based on synthesis feedback. In this paper, we present an automatic RTL code optimization framework leveraging large models. We introduce several key components to address the challenges, including a circuit partition pipeline, large model-aided RTL optimization encompassing multi-modal program analysis, a search engine, and a cost-aware search algorithm for efficient rewriting. Additionally, we have developed a fast verification pipeline to streamline the verification process and reduce costs. Moreover, we have created two datasets to foster further advancements in RTL code optimization. We hope our work can stimulate innovation in RTL code optimization.

\clearpage
{
    \bibliographystyle{IEEEtran}
    \bibliography{ref/RTL-opt} 

\begin{thebibliography}{10}
\providecommand{\url}[1]{#1}
\csname url@samestyle\endcsname
\providecommand{\newblock}{\relax}
\providecommand{\bibinfo}[2]{#2}
\providecommand{\BIBentrySTDinterwordspacing}{\spaceskip=0pt\relax}
\providecommand{\BIBentryALTinterwordstretchfactor}{4}
\providecommand{\BIBentryALTinterwordspacing}{\spaceskip=\fontdimen2\font plus
\BIBentryALTinterwordstretchfactor\fontdimen3\font minus
  \fontdimen4\font\relax}
\providecommand{\BIBforeignlanguage}[2]{{%
\expandafter\ifx\csname l@#1\endcsname\relax
\typeout{** WARNING: IEEEtran.bst: No hyphenation pattern has been}%
\typeout{** loaded for the language `#1'. Using the pattern for}%
\typeout{** the default language instead.}%
\else
\language=\csname l@#1\endcsname
\fi
#2}}
\providecommand{\BIBdecl}{\relax}
\BIBdecl

\bibitem{wolf2013yosys}
C.~Wolf, J.~Glaser, and J.~Kepler, ``Yosys-a free verilog synthesis suite,'' in
  \emph{Proceedings of the 21st Austrian Workshop on Microelectronics
  (Austrochip)}, vol.~97, 2013.

\bibitem{pasko1999new}
R.~Pasko, P.~Schaumont, V.~Derudder, S.~Vernalde, and D.~Durackova, ``A new
  algorithm for elimination of common subexpressions,'' \emph{IEEE Transactions
  on Computer-Aided Design of Integrated Circuits and Systems}, vol.~18, no.~1,
  pp. 58--68, 1999.

\bibitem{wegman1991constant}
M.~N. Wegman and F.~K. Zadeck, ``Constant propagation with conditional
  branches,'' \emph{ACM Transactions on Programming Languages and Systems
  (TOPLAS)}, vol.~13, no.~2, pp. 181--210, 1991.

\bibitem{chen2004register}
D.~Chen and J.~Cong, ``Register binding and port assignment for multiplexer
  optimization,'' in \emph{ASP-DAC 2004: Asia and South Pacific Design
  Automation Conference 2004 (IEEE Cat. No. 04EX753)}.\hskip 1em plus 0.5em
  minus 0.4em\relax IEEE, 2004, pp. 68--73.

\bibitem{pivstekareduction}
P.~Pi{\v{s}}teka, K.~Jelemensk{\'a}, and M.~Koles{\'a}r, ``Reduction of
  multiplexer trees using modified lookup table.''

\bibitem{wang2023optimization}
Z.~Wang, H.~You, J.~Wang, M.~Liu, Y.~Su, and Y.~Zhang, ``Optimization of
  multiplexer combination in rtl logic synthesis,'' in \emph{2023 International
  Symposium of Electronics Design Automation (ISEDA)}.\hskip 1em plus 0.5em
  minus 0.4em\relax IEEE, 2023, pp. 121--125.

\bibitem{cocke1970global}
J.~Cocke, ``Global common subexpression elimination,'' in \emph{Proceedings of
  a symposium on Compiler optimization}, 1970, pp. 20--24.

\bibitem{metzger1993interprocedural}
R.~Metzger and S.~Stroud, ``Interprocedural constant propagation: An empirical
  study,'' \emph{ACM Letters on Programming Languages and Systems (LOPLAS)},
  vol.~2, no. 1-4, pp. 213--232, 1993.

\bibitem{buchberger1982algebraic}
B.~Buchberger and R.~Loos, ``Algebraic simplification,'' in \emph{Computer
  algebra: symbolic and algebraic computation}.\hskip 1em plus 0.5em minus
  0.4em\relax Springer, 1982, pp. 11--43.

\bibitem{carette2004understanding}
J.~Carette, ``Understanding expression simplification,'' in \emph{Proceedings
  of the 2004 international symposium on Symbolic and algebraic computation},
  2004, pp. 72--79.

\bibitem{knoop1994partial}
J.~Knoop, O.~R{\"u}thing, and B.~Steffen, ``Partial dead code elimination,''
  \emph{ACM Sigplan Notices}, vol.~29, no.~6, pp. 147--158, 1994.

\bibitem{gupta1997path}
R.~Gupta, D.~Benson, and J.~Z. Fang, ``Path profile guided partial dead code
  elimination using predication,'' in \emph{Proceedings 1997 International
  Conference on Parallel Architectures and Compilation Techniques}.\hskip 1em
  plus 0.5em minus 0.4em\relax IEEE, 1997, pp. 102--113.

\bibitem{cooper2001operator}
K.~D. Cooper, L.~T. Simpson, and C.~A. Vick, ``Operator strength reduction,''
  \emph{ACM Transactions on Programming Languages and Systems (TOPLAS)},
  vol.~23, no.~5, pp. 603--625, 2001.

\bibitem{laforest2010efficient}
C.~E. LaForest and J.~G. Steffan, ``Efficient multi-ported memories for
  fpgas,'' in \emph{Proceedings of the 18th annual ACM/SIGDA international
  symposium on Field programmable gate arrays}, 2010, pp. 41--50.

\bibitem{ma2020hypervisor}
J.~Ma, G.~Zuo, K.~Loughlin, X.~Cheng, Y.~Liu, A.~M. Eneyew, Z.~Qi, and
  B.~Kasikci, ``A hypervisor for shared-memory fpga platforms,'' in
  \emph{Proceedings of the Twenty-Fifth International Conference on
  Architectural Support for Programming Languages and Operating Systems}, 2020,
  pp. 827--844.

\bibitem{zhou2017new}
Y.~Zhou, K.~M. Al-Hawaj, and Z.~Zhang, ``A new approach to automatic memory
  banking using trace-based address mining,'' in \emph{Proceedings of the 2017
  ACM/SIGDA International Symposium on Field-Programmable Gate Arrays}, 2017,
  pp. 179--188.

\bibitem{lai2019remap+}
B.-C. Lai, B.-Y. Chen, B.-E. Chen, and Y.-D. Hsin, ``Remap+: An efficient
  banking architecture for multiple writes of algorithmic memory,'' \emph{IEEE
  Transactions on Very Large Scale Integration (VLSI) Systems}, vol.~28, no.~3,
  pp. 660--671, 2019.

\bibitem{park2001synthesis}
J.~Park and P.~C. Diniz, ``Synthesis of pipelined memory access controllers for
  streamed data applications on fpga-based computing engines,'' in
  \emph{Proceedings of the 14th international symposium on Systems synthesis},
  2001, pp. 221--226.

\bibitem{kam2013synthesis}
T.~Kam, T.~Villa, R.~K. Brayton, and A.~L. Sangiovanni-Vincentelli,
  \emph{Synthesis of finite state machines: functional optimization}.\hskip 1em
  plus 0.5em minus 0.4em\relax Springer Science \& Business Media, 2013.

\bibitem{villa2012synthesis}
T.~Villa, T.~Kam, R.~K. Brayton, and A.~L. Sangiovanni-Vincentelli,
  \emph{Synthesis of finite state machines: logic optimization}.\hskip 1em plus
  0.5em minus 0.4em\relax Springer Science \& Business Media, 2012.

\bibitem{shelar1999decomposition}
R.~S. Shelar, M.~P. Desai, and H.~Narayanan, ``Decomposition of finite state
  machines for area, delay minimization,'' in \emph{Proceedings 1999 IEEE
  International Conference on Computer Design: VLSI in Computers and Processors
  (Cat. No. 99CB37040)}.\hskip 1em plus 0.5em minus 0.4em\relax IEEE, 1999, pp.
  620--625.

\bibitem{liu2023chipnemo}
M.~Liu, T.-D. Ene, R.~Kirby, C.~Cheng, N.~Pinckney, R.~Liang, J.~Alben,
  H.~Anand, S.~Banerjee, I.~Bayraktaroglu \emph{et~al.}, ``Chipnemo:
  Domain-adapted llms for chip design,'' \emph{arXiv preprint
  arXiv:2311.00176}, 2023.

\bibitem{blocklove2023chip}
J.~Blocklove, S.~Garg, R.~Karri, and H.~Pearce, ``Chip-chat: Challenges and
  opportunities in conversational hardware design,'' in \emph{2023 ACM/IEEE 5th
  Workshop on Machine Learning for CAD (MLCAD)}.\hskip 1em plus 0.5em minus
  0.4em\relax IEEE, 2023, pp. 1--6.

\bibitem{fu2023gpt4aigchip}
Y.~Fu, Y.~Zhang, Z.~Yu, S.~Li, Z.~Ye, C.~Li, C.~Wan, and Y.~C. Lin,
  ``Gpt4aigchip: Towards next-generation ai accelerator design automation via
  large language models,'' in \emph{2023 IEEE/ACM International Conference on
  Computer Aided Design (ICCAD)}.\hskip 1em plus 0.5em minus 0.4em\relax IEEE,
  2023, pp. 1--9.

\bibitem{zhang2024data4aigchip}
Y.~Zhang, Y.~Fu, Z.~Yu, K.~Zhao, C.~Wan, C.~Li, and Y.~C. Lin, ``Data4aigchip:
  An automated data generation and validation flow for llm-assisted hardware
  design,'' 2024.

\bibitem{wu2024chateda}
H.~Wu, Z.~He, X.~Zhang, X.~Yao, S.~Zheng, H.~Zheng, and B.~Yu, ``Chateda: A
  large language model powered autonomous agent for eda,'' \emph{IEEE
  Transactions on Computer-Aided Design of Integrated Circuits and Systems},
  2024.

\bibitem{thakur2023verigen}
S.~Thakur, B.~Ahmad, H.~Pearce, B.~Tan, B.~Dolan-Gavitt, R.~Karri, and S.~Garg,
  ``Verigen: A large language model for verilog code generation,'' \emph{arXiv
  preprint arXiv:2308.00708}, 2023.

\bibitem{liu2023rtlcoder}
S.~Liu, W.~Fang, Y.~Lu, Q.~Zhang, H.~Zhang, and Z.~Xie, ``Rtlcoder:
  Outperforming gpt-3.5 in design rtl generation with our open-source dataset
  and lightweight solution,'' \emph{arXiv preprint arXiv:2312.08617}, 2023.

\bibitem{delorenzo2024make}
M.~DeLorenzo, A.~B. Chowdhury, V.~Gohil, S.~Thakur, R.~Karri, S.~Garg, and
  J.~Rajendran, ``Make every move count: Llm-based high-quality rtl code
  generation using mcts,'' \emph{arXiv preprint arXiv:2402.03289}, 2024.

\bibitem{tsai2023rtlfixer}
Y.~Tsai, M.~Liu, and H.~Ren, ``Rtlfixer: Automatically fixing rtl syntax errors
  with large language models,'' \emph{arXiv preprint arXiv:2311.16543}, 2023.

\bibitem{pei2024betterv}
Z.~Pei, H.-L. Zhen, M.~Yuan, Y.~Huang, and B.~Yu, ``Betterv: Controlled verilog
  generation with discriminative guidance,'' \emph{arXiv preprint
  arXiv:2402.03375}, 2024.

\bibitem{yao2024hdldebugger}
X.~Yao, H.~Li, T.~H. Chan, W.~Xiao, M.~Yuan, Y.~Huang, L.~Chen, and B.~Yu,
  ``Hdldebugger: Streamlining hdl debugging with large language models,''
  \emph{arXiv preprint arXiv:2403.11671}, 2024.

\bibitem{wang2023circuit}
Z.~Wang, L.~Chen, J.~Wang, X.~Li, Y.~Bai, X.~Li, M.~Yuan, J.~Hao, Y.~Zhang, and
  F.~Wu, ``A circuit domain generalization framework for efficient logic
  synthesis in chip design,'' \emph{arXiv preprint arXiv:2309.03208}, 2023.

\bibitem{wang2024hierarchical}
Z.~Wang, J.~Wang, D.~Zuo, J.~Yunjie, X.~Xia, Y.~Ma, H.~Jianye, M.~Yuan,
  Y.~Zhang, and F.~Wu, ``A hierarchical adaptive multi-task reinforcement
  learning framework for multiplier circuit design,'' in \emph{Forty-first
  International Conference on Machine Learning}, 2024.

\bibitem{liu2024unified}
T.~Liu, Y.~Sun, L.~Chen, X.~Li, M.~Yuan, and E.~F. Young, ``A unified parallel
  framework for lut mapping and logic optimization,'' \emph{IEEE Transactions
  on Computer-Aided Design of Integrated Circuits and Systems}, 2024.

\bibitem{li2024logic}
X.~Li, X.~Li, L.~Chen, X.~Zhang, M.~Yuan, and J.~Wang, ``Logic synthesis with
  generative deep neural networks,'' \emph{arXiv preprint arXiv:2406.04699},
  2024.

\bibitem{li2024circuit}
------, ``Circuit transformer: End-to-end circuit design by predicting the next
  gate,'' \emph{arXiv preprint arXiv:2403.13838}, 2024.

\bibitem{liu2023verilogeval}
M.~Liu, N.~Pinckney, B.~Khailany, and H.~Ren, ``Verilogeval: Evaluating large
  language models for verilog code generation,'' in \emph{2023 IEEE/ACM
  International Conference on Computer Aided Design (ICCAD)}.\hskip 1em plus
  0.5em minus 0.4em\relax IEEE, 2023, pp. 1--8.

\bibitem{liu2024lost}
N.~F. Liu, K.~Lin, J.~Hewitt, A.~Paranjape, M.~Bevilacqua, F.~Petroni, and
  P.~Liang, ``Lost in the middle: How language models use long contexts,''
  \emph{Transactions of the Association for Computational Linguistics},
  vol.~12, pp. 157--173, 2024.

\bibitem{chen2016xgboost}
T.~Chen and C.~Guestrin, ``Xgboost: A scalable tree boosting system,'' 2016,
  pp. 785--794.

\bibitem{wei2022chain}
J.~Wei, X.~Wang, D.~Schuurmans, M.~Bosma, F.~Xia, E.~Chi, Q.~V. Le, D.~Zhou
  \emph{et~al.}, ``Chain-of-thought prompting elicits reasoning in large
  language models,'' \emph{Advances in Neural Information Processing Systems},
  vol.~35, pp. 24\,824--24\,837, 2022.

\bibitem{gao2023retrieval}
Y.~Gao, Y.~Xiong, X.~Gao, K.~Jia, J.~Pan, Y.~Bi, Y.~Dai, J.~Sun, and H.~Wang,
  ``Retrieval-augmented generation for large language models: A survey,''
  \emph{arXiv preprint arXiv:2312.10997}, 2023.

\bibitem{ioannidis1996query}
Y.~E. Ioannidis, ``Query optimization,'' \emph{ACM Computing Surveys (CSUR)},
  vol.~28, no.~1, pp. 121--123, 1996.

\bibitem{aizawa2003information}
A.~Aizawa, ``An information-theoretic perspective of tf--idf measures,''
  \emph{Information Processing \& Management}, vol.~39, no.~1, pp. 45--65,
  2003.

\bibitem{auer2002using}
P.~Auer, ``Using confidence bounds for exploitation-exploration trade-offs,''
  \emph{Journal of Machine Learning Research}, vol.~3, no. Nov, pp. 397--422,
  2002.

\bibitem{lavagno2016electronic}
L.~Lavagno, I.~L. Markov, G.~Martin, and L.~K. Scheffer, \emph{Electronic
  design automation for IC implementation, circuit design, and process
  technology: circuit design, and process technology}.\hskip 1em plus 0.5em
  minus 0.4em\relax CRC Press, 2016.

\bibitem{marques1999combinational}
J.~Marques-Silva and T.~Glass, ``Combinational equivalence checking using
  satisfiability and recursive learning,'' in \emph{Proceedings of the
  conference on Design, automation and test in Europe}, 1999, pp. 33--es.

\bibitem{sayed2016equivalence}
A.~Sayed-Ahmed, D.~Gro{\ss}e, M.~Soeken, and R.~Drechsler, ``Equivalence
  checking using gr{\"o}bner bases,'' in \emph{2016 Formal Methods in
  Computer-Aided Design (FMCAD)}.\hskip 1em plus 0.5em minus 0.4em\relax IEEE,
  2016, pp. 169--176.

\bibitem{godefroid2008automated}
P.~Godefroid, M.~Y. Levin, D.~A. Molnar \emph{et~al.}, ``Automated whitebox
  fuzz testing.'' in \emph{NDSS}, vol.~8, 2008, pp. 151--166.

\bibitem{achiam2023gpt}
J.~Achiam, S.~Adler, S.~Agarwal, L.~Ahmad, I.~Akkaya, Aleman \emph{et~al.},
  ``Gpt-4 technical report,'' \emph{arXiv preprint arXiv:2303.08774}, 2023.

\bibitem{coward2022automatic}
S.~Coward, G.~A. Constantinides, and T.~Drane, ``Automatic datapath
  optimization using e-graphs,'' in \emph{2022 IEEE 29th Symposium on Computer
  Arithmetic (ARITH)}.\hskip 1em plus 0.5em minus 0.4em\relax IEEE, 2022, pp.
  43--50.

\bibitem{dosovitskiy2020image}
A.~Dosovitskiy, L.~Beyer, A.~Kolesnikov, D.~Weissenborn, X.~Zhai,
  T.~Unterthiner, M.~Dehghani, M.~Minderer, G.~Heigold, S.~Gelly \emph{et~al.},
  ``An image is worth 16x16 words: Transformers for image recognition at
  scale,'' \emph{arXiv preprint arXiv:2010.11929}, 2020.

\bibitem{touvron2023llama}
H.~Touvron, L.~Martin, K.~Stone, P.~Albert, A.~Almahairi, Y.~Babaei,
  N.~Bashlykov, S.~Batra, P.~Bhargava, S.~Bhosale \emph{et~al.}, ``Llama 2:
  Open foundation and fine-tuned chat models,'' \emph{arXiv preprint
  arXiv:2307.09288}, 2023.

\bibitem{bi2024deepseek}
X.~Bi, D.~Chen, G.~Chen, S.~Chen, D.~Dai \emph{et~al.}, ``Deepseek llm: Scaling
  open-source language models with longtermism,'' \emph{arXiv preprint
  arXiv:2401.02954}, 2024.

\bibitem{williams2002icarus}
S.~Williams and M.~Baxter, ``Icarus verilog: open-source verilog more than a
  year later,'' \emph{Linux Journal}, vol. 2002, no.~99, p.~3, 2002.

\bibitem{brayton2010abc}
R.~Brayton and A.~Mishchenko, ``Abc: An academic industrial-strength
  verification tool,'' in \emph{Computer Aided Verification: 22nd International
  Conference, CAV 2010, Edinburgh, UK, July 15-19, 2010. Proceedings 22}.\hskip
  1em plus 0.5em minus 0.4em\relax Springer, 2010, pp. 24--40.

\bibitem{robertson2009probabilistic}
S.~Robertson, H.~Zaragoza \emph{et~al.}, ``The probabilistic relevance
  framework: Bm25 and beyond,'' \emph{Foundations and Trends{\textregistered}
  in Information Retrieval}, vol.~3, no.~4, pp. 333--389, 2009.

\bibitem{liaw2002classification}
A.~Liaw, M.~Wiener \emph{et~al.}, ``Classification and regression by
  randomforest,'' \emph{R news}, vol.~2, no.~3, pp. 18--22, 2002.

\end{thebibliography}
}


\end{document}